\pdfoutput=1

\documentclass[aps]{revtex4}

\usepackage{graphicx,subfigure}
\usepackage{amsmath,amssymb}
\usepackage{bm}

\newcommand{\be}{\begin{eqnarray}}
\newcommand{\ee}{\end{eqnarray}}
\renewcommand{\sc}{\slashchar}
\def\slashchar#1{\setbox0=\hbox{$#1$}           
   \dimen0=\wd0                                 
   \setbox1=\hbox{/} \dimen1=\wd1               
  \ifdim\dimen0>\dimen1                        
 \rlap{\hbox to \dimen0{\hfil/\hfil}}      
  #1                                        
 \else                                        
    \rlap{\hbox to \dimen1{\hfil$#1$\hfil}}   
    /                                         
 \fi}                                         %

\begin{document}

\title{Instanton fermionic zero mode at finite temperature and chemical potential}

\author{ Marco Cristoforetti}
\affiliation{E.C.T.$^\star$, Strada delle Tabarelle 286, Villazzano (Trento), I-38123 Italy}
\affiliation{LISC, Via Sommarive 18, Povo (Trento), I-38123  Italy}

\email{mcristofo@ectstar.eu}
\begin{abstract}
The spontaneous breaking of chiral symmetry and the U(1) axial anomaly in QCD, can be phenomenologically understood by considering instantons as the gauge configurations mediating quark-quark interaction. The existence of an exact zero mode solution of the Dirac equation in the field of a single instanton is the fundamental ingredient. Explicit expressions for $\psi_0$ are available for $T\neq 0$ and $\mu=0$, and $\mu\neq0$ and $T=0$. In this paper we derive the solution for the most general case $T\neq0$ and $\mu\neq0$. This opens the possibility of investigating the QCD dynamics associated with instantons in the full phase diagram. As a first step in this direction, we will study the dependence of the instanton density from the thermodynamic coordinates.
\end{abstract}
\maketitle

\section{Introduction}

Our present understanding of the strong interaction is limited by the lack of analytical tools to investigate QCD in the infrared regime, where the theory is non-perturbative. Presently, the only rigorous non-perturbative framework to solve QCD is represented by ab-initio lattice simulations; however, in this framework, it is difficult to gain sufficient insight about the nature of the gauge field fluctuations which dominate the dynamics. This is essentially due to the method generally adopted for generating the statistical ensemble, since Monte Carlo importance sampling algorithms does not yield the configuration details needed for calculating the expectation values of interest.

To shed some light on the mechanisms that underlie the QCD non-perturbative regime, a plethora of different models have been developed, in which the dynamical effects obtained can be clearly connected with the presence of specific degrees of freedom in the vacuum. The instanton model is a paradigmatic example of such models.

The instanton is a well known semiclassical gauge configuration with integer topological charge, corresponding, in Euclidean space, to an exact self-dual solution of the Yang-Mills equations of motion. In the seventies a deep connection was established between QCD instantons and important properties of the theory, like chiral symmetry breaking and the $U(1)$ axial anomaly~\cite{'tHooft:1976up,'tHooft:1976fv}. It turns out that both phenomena are strongly related to the existence of zero modes and near-zero modes of the Dirac operator. In particular the violation of the axial charge in an arbitrary gauge potential is given by $\Delta q_5=2 N_f(n_L-n_R)$, where $n_{L,R}$ is the number of left and right-handed zero modes of the Dirac operator. 

Since an (anti-)instanton has one and only one exact (right-)left-handed zero mode, the anomaly emerges naturally at the level of the single instanton calculation. On the other hand, chiral symmetry breaking can be related to a finite density of quasi-zero mode through the Bank-Casher formula~\cite{Banks:1979yr} $\langle\overline{q}q\rangle=-\pi\rho(\lambda=0)$, where $\rho$ is the density of modes  with eigenvalue $\lambda$. The distribution of near-zero mode predicted by the Instanton Liquid Model, where a dilute and globally topologically neutral ensemble of pseudoparticle is considered, perfectly agrees with the phenomenological expectation and with calculations in Chiral Perturbation Theory~\cite{Schafer:1996wv,Cristoforetti:2006ar,Cristoforetti:2007ak}. 

More in general, the model provides a successful framework in the phenomenological study of the chiral dynamics of QCD, as proved extensively in literature~\cite{Schafer:1995pz,Schafer:1995uz,Faccioli:2003qz,Cristoforetti:2006ar,Cristoforetti:2007ak,Cristoforetti:2004kj,Cristoforetti:2004rr}. 

In this work we will investigate the QCD phase diagram in the instanton model. For achieving this, one needs to extend the latter model to allow the study of the thermodynamic quantities. This requires, not only the analytic form of the instanton field at finite temperature ($T$)~\cite{Harrington:1978ve}, but also the general expression for the zero mode at finite temperature and chemical potential ($\mu$). To date, the zero mode is known analytically only for the case of finite temperature with vanishing quark chemical potential and finite quark chemical potential with zero temperature~\cite{AragaodeCarvalho:1980de,Ilgenfritz:1988dh,Ilgenfritz:1994nt,Velkovsky:1996aj,Schafer:1998up,Rapp:1999qa,Schafer:2002ty}; as a consequence the phenomenology of the instanton model has not been extended to the full phase diagram. One attempt in this direction was pursued by Shuryak and collaborators~\cite{Rapp:1999qa} who, by extrapolating across the full phase diagram the results obtained along the axis, tried to extract general information on the chiral symmetry restoration. Nevertheless the absence of an analytic expression for the zero mode in the case of both finite $T$ and finite $\mu$ prevented to clarify in an unambiguous way the prediction of the model.

In this paper we present the analytic expression of the zero mode as obtained through solving the Dirac equation in the background of a single instanton for the general case of $T\neq0$ and $\mu_q\neq0$. With this new ingredient, all the terms entering the partition function of the model are known, and the study of the phase diagram of the instanton ensemble can be in principle carried out. As a first step we compute here the dependency of the instanton density  from the temperature and chemical potential. The relevance of this last analysis can be understood looking at the relation $\langle q\bar{q}\rangle~=~-1/(\pi\rho)(3n/2)^{1/2}$ which connect the order parameter of chiral symmetry breaking $\langle q\bar{q}\rangle$, with a finite instanton density in the vacuum $n$.

The paper is organized as follows. Sec.~\ref{sec:zeromode} is devoted to the analysis of instanton induced fermionic zero modes. A review of the main features of these modes is presented in Sec.~\ref{sec:chirality}, where we spend some time focusing on the property of fixed chirality of the zero modes. Next, in Sec.~\ref{sec:zmodemut} we present the main result of the paper: starting from an Ansatz for the fermionic field profile, we solve the Dirac equation in the single instanton background at finite temperature and chemical potential, showing that the proposed profile is indeed a zero mode.Next, in Sec.~\ref{sec:phasediagram}, we use the result obtained in the first part to start proving a systematic study of the phase diagram of the (instanton) model. To begin with in Sec.~\ref{sec:partfunctia} we give an explanation of the different terms entering the partition function of the model, to show how the knowledge of the analytic expression of the zero mode makes it possible to study the system for any $T$ and $\mu$.  In Sec.~\ref{sec:chiraltrans} we then show the instanton density in the full phase diagram and connect it to chiral symmetry breaking. Finally we present our conclusions and outlooks in Sec.~\ref{sec:conclusion} .

\section{Fermionic zero modes}\label{sec:zeromode}
Instantons are self-dual classical solution of the Yang-Mills Euclidean equation of motions with fixed integer topological charge. In this paper we concentrate on singular gauge instanton, though all the results can be simply mapped to the regular gauge case. 

The instanton field is written in the form
\be\label{eq:amu}
	A_{\mu}^a(x)=-\bar{\eta}^a_{\mu\nu}\partial_{\nu}\ln\Pi(x),
\ee
where $\bar{\eta}_{\mu\nu}^a$ is the 't Hooft symbol (main properties of it being presented in Appendix~\ref{app:thoofts}) while the choice of the profile function $\Pi(x)$ is dictated by the requirement of self-duality of the field strength $F_{\mu\nu}^a(x)=\tilde{F}_{\mu\nu}^a(x)$. Indeed, substituting (\ref{eq:amu}) in the definition of $F_{\mu\nu}^a$ one has
\be\label{eq:fmunu}
	F_{\mu\nu}^a(x)&=&\bar{\eta}^a_{\nu\rho}(-\partial_{\mu}\partial_{\rho}\ln\Pi(x)+(\partial_{\mu}\ln\Pi(x))(\partial_{\rho}\ln\Pi(x)))\nonumber\\
	&&-(\mu\leftrightarrow\nu)+\bar{\eta}^a_{\mu\nu}(\partial_{\rho}\ln\Pi(x))^2,
\ee
and self-duality is guaranteed only if it is such that
\be\label{eq:selfdual}
	\Box\ln\Pi(x)+(\partial_{\rho}\ln\Pi(x))^2=0.
\ee
The choice of $\Pi(x)=1+\rho^2/x^2$ satisfies the condition (\ref{eq:selfdual}) and the parameter $\rho$ identifies the size of the instanton.

\subsection{Chirality of the zero mode}\label{sec:chirality}

This paper is mainly devoted to the study of the zero mode of the Dirac equation as induced by instantons, at finite temperature and chemical potential. In this section we start the analysis reviewing the proof of one of the main characteristic of this fermionic modes: zero modes, associated with the presence of single instantons in the QCD vacuum, have fixed chirality. This information will be helpful in the following, searching for the analytic form of the zero mode.

An important and direct consequence of this property lies in the very specific form of the instanton mediate quark interaction: two quarks exchanging an instanton necessarily have to flip their chirality. This effect is a very peculiar feature of the instanton gauge field, and has allowed to prove the presence of a finite density of instanton in the ground state of the theory~\cite{Faccioli:2003qz}. 

By definition a fermionic field $\psi_0(x)$ is a zero mode if 
\be\label{eq:direq}
	i\sc{D}\psi(x)=i\gamma_{\mu}(\partial^{\mu}-iA_I^{\mu}(x))\psi(x)=0,
\ee
where $A_I^{\mu}$ is the instanton profile previously introduced. To prove that such a zeromode would have fixed chirality, it is useful to start the analysis from the square of Dirac operator. Eq.~\ref{eq:direq} can be squared yielding the equation $\sc{D}^2\psi(x)=0$ where the square of the Dirac operator $\sc{D}^2$ is given by
\be
	(\gamma_{\mu}D^{\mu})^2=D^2+\dfrac{i}{4}[\gamma_{\mu},\gamma_{\nu}]F_{\mu\nu}(x).
\ee
At this point the adoption of a quaternionic notation, presented in Appendix~\ref{app:notation}, results particularly convenient. In this context the commutator of the gamma matrices can be expressed in terms of the 't Hooft symbol and $\sigma_\mu$ matrices as
\begin{displaymath}
[\gamma_{\mu},\gamma_{\nu}]=\gamma_{\mu}\gamma_{\nu}-\gamma_{\nu}\gamma_{\mu}=2i
	\left(\begin{array}{cc}
		\bar{\eta}^a_{\mu\nu}\sigma^a & 0\\
		0&\eta^a_{\mu\nu}\sigma^a
		\end{array}
	\right).
\end{displaymath}
Throughout the paper we discuss the case of massless quarks, where the Dirac spinor can be written as a sum of a right and left handed mode: $\psi=\psi_R+\psi_L$. This means that for the definition of the squared Dirac equation we have
\be\label{eq:dsq}
\Big(D^2+\dfrac{i}{4}[\gamma_{\mu},\gamma_{\nu}]F^{\mu\nu}\Big)\psi=
	\left(\begin{array}{cc}
		D^2-\dfrac{\sigma^a}{2}\bar{\eta}^a_{\mu\nu}F^{\mu\nu} & 0\\
		0&D^2-\dfrac{\sigma^a}{2}\eta^a_{\mu\nu}F^{\mu\nu}
		\end{array}
	\right)\left(\begin{array}{c}
		\vspace{.5cm}\psi_R\\
		\vspace{.1cm}\psi_L
		\end{array}
	\right)=0.
\ee
Considering the field strength (\ref{eq:fmunu}) and using the properties of the 't Hooft symbol (Appendix~\ref{app:thoofts}), the term $\sigma^a\bar{\eta}^a_{\mu\nu}F^{\mu\nu}$, which relates to the right handed component of the spinor, can be rewritten in a more compact form as 
\be
	\dfrac{\sigma^a}{2}\bar{\eta}^a_{\mu\nu}F^{\mu\nu}&=&\dfrac{\sigma^a}{2}\bar{\eta}^a_{\mu\nu}[\bar{\eta}^a_{\nu\rho}(-\partial_{\mu}\partial_{\rho}\ln\Pi+(\partial_{\mu}\ln\Pi)(\partial_{\rho}\ln\Pi))-(\mu\leftrightarrow\nu)+
		\bar{\eta}^a_{\mu\nu}(\partial_{\rho}\ln\Pi)^2]\nonumber\\
		&=&\dfrac{3}{2}\left[\Box\ln\Pi+(\partial_{\rho}\ln\Pi)^2\right].
\ee

Here we recognize the left hand side of the self-duality condition (\ref{eq:selfdual}); since for an instanton $\Box\ln\Pi+(\partial_{\rho}\ln\Pi)^2=0$, we have that for the right component of the quark field the equation
\be\label{eq:rzm}
	D^2\psi_R=0,
\ee
holds. Moreover, since  $D$ is an hermitian operator, its square is positive definite, so that the equation (\ref{eq:rzm}) has no solution. The conclusion is that, if the zero mode exists, it must be left handed and such that it satisfies the equation
\be
	\left(D^2-\dfrac{\sigma^a}{2}\eta^a_{\mu\nu}F_{\mu\nu}\right)\psi_L=0
\ee
The fixed chirality of the zero mode will help us in the research of the analytic expression for the zero mode.

\subsection{Zero mode profile at $T\neq0$ and $\mu\neq0$}\label{sec:zmodemut}
In this section we present our main result, namely we derive the quark zero mode profile in the general case of non-vanishing temperature and chemical potential. The starting point is the expression of the zero mode for the two limiting case $T=0$, $\mu\neq0$ and $T\neq0$, $\mu=0$~\cite{Schafer:1996wv}:
\be\label{eq:ans}
	\psi_0(r,t)=\dfrac{1}{2\pi\rho}e^{\mu t}\sqrt{\Pi(r,t)}\sc{\partial}\left(\dfrac{\Phi(r,t)}{\Pi(r,t)}e^{-\mu t}\right)\gamma_{\pm}\varepsilon
\ee
where $\gamma_{\pm}$ projects on the desired chirality, and 
\begin{displaymath}
[\varepsilon]_{\mu a}=\dfrac{1}{\sqrt{2}}
	\left(\begin{array}{c}
		-i\sigma_2\\
		i\sigma_2
		\end{array}
	\right)_{\mu a}
\end{displaymath}
is a Dirac and color spinor. The other two functions entering expression (\ref{eq:ans}) are the $\Pi(x)$ instanton profile, which here differentiates in the case of zero and finite temperature:
 \be\label{eq:Pi}
	\Pi(r,t)=
	\left\{\begin{array}{cc}
	1+\dfrac{\rho^2}{r^2+t^2} & T=0\\ &\\
	1+\dfrac{\pi\rho^2}{T r}\dfrac{\sinh(2\pi r T)}{\cosh(2\pi r T)-\cos(2\pi t T)} & T>0
	\end{array}\right. ,
\ee
and the new profile function $\Phi(r,t)$ defined by
\be
	\Phi(r,t)=\left(\Pi(r,t)-1\right)\left\{\begin{array}{cc}
		\cos(r\mu)+\dfrac{t}{r}\sin(r \mu) & \mu\neq 0,\ T=0 \\ &\\
		\dfrac{\cos(\pi t T)}{\cosh(\pi r T)} & \mu=0,\ T\neq 0
	\end{array}\right. .
\ee

The rest of this section is devoted to prove that, in order to obtain the generalized zero mode at non vanishing $T$ and $\mu$, we only have to replace the $\Phi(r,t)$ above with the Ansatz
\be\label{eq:phi}
	\Phi(r,t)=\left(\Pi(r,t)-1\right)\left(\dfrac{\cos(\pi t T)\cos(r \mu)}{\cosh(\pi r T)}+\dfrac{\sin(\pi t T)\sin(r \mu)}{\sinh(\pi r T)}\right)
\ee

To achieve our goal we proceed by inspection, proving that (\ref{eq:phi}) is indeed a zero mode solution, i.e. that:
\be\label{eq:deq}
	i(\sc{\partial}-i\sc{A}^I-\mu\gamma_4)\psi(x;\mu,T)=0.
\ee
It is important to notice here that, in the case of finite chemical potential, the solution of the adjoint Dirac equation,
\be
	\psi^\dag(x;-\mu,T)i(\sc{\partial}-i\sc{A}^I-\mu\gamma_4)=0.	
\ee 
carries the chemical potential argument with opposite sign. This is crucial in order to have a consistent definition of the expectation values at finite $\mu$. In particular looking at the case of zero temperature this property renders finite the norm,
\be
	\int\textrm{d}^4x\psi^\dag(x;-\mu)\psi^\dag(x;\mu)=1,
\ee
whereas without the extra sign one has
\be
	\int\textrm{d}^4x\psi^\dag(x;\mu)\psi^\dag(x;\mu)=\infty.
\ee
This singularity being connected with the BCS singularity one encounters when resumming an effective (attractive) particle-particle interaction around the Fermi surface.

To begin with we pass to the quaternionic formalism and write the zero mode profile as: 
\be\nonumber
	[\psi_0]_{j\beta}&=&\Bigg[\dfrac{e^{\mu t} \sqrt{\Pi}}{2\pi\sqrt{2}\rho}
	\left(\begin{array}{cc}
		0 & -i\sigma_\rho^{\dag}\\
		i\sigma_\rho&0
		\end{array}
	\right)_{\beta\gamma}
	\left(\begin{array}{c}
		-i\sigma_2\\
		i\sigma2
		\end{array}
	\right)_{\gamma j}
	\partial_\rho\left(\dfrac{\Phi}{\Pi}e^{-\mu t}\right)
	\Bigg]^T\\\nonumber
	&=& \left[\dfrac{e^{\mu t} \sqrt{\Pi}}{2\pi\sqrt{2}\rho}
	\left(\begin{array}{c}
		-i\sigma_\rho^{\dag} i\sigma_2\\
		-i\sigma_\rho i\sigma_2
		\end{array}
	\right)_{\beta j}
	\partial_\rho\left(\dfrac{\Phi}{\Pi}e^{-\mu t}\right)
	\right]^T\\
	\nonumber
	&=&\dfrac{e^{\mu t} \sqrt{\Pi}}{2\pi\sqrt{2}\rho}
	\left[\begin{array}{cc}
		\sigma_2^T \sigma_\rho^{\dag T} & \sigma_2^T \sigma_\rho^T
		\end{array}
	\right]_{j\beta}
	\partial_\rho\left(\dfrac{\Phi}{\Pi}e^{-\mu t}\right).
\ee

Making then use of the properties of the Pauli matrices
\be
	\sigma_2^T=-\sigma_2 &\ \textrm{and}&\ \sigma_\rho^T=-\sigma_2\sigma_\rho^\dag\sigma_2,
\ee
we obtain the explicit expression for the color ($j$) and spinor $(\beta)$ components of the zero mode, namely
\be\label{eq:zmodequat}
	[\psi_0]_{j\beta}	&=&\dfrac{e^{\mu t} \sqrt{\Pi}}{2\pi\sqrt{2}\rho}
	\left[\begin{array}{cc}
		\sigma_\rho\sigma_2 & \sigma_\rho^\dag\sigma_2
		\end{array}
	\right]_{j\beta}
	\partial_\rho\left(\dfrac{\Phi}{\Pi}e^{-\mu t}\right).
\ee

At this point, we can use the fact that the zero mode is left-handed; projecting on the left-hand component of~(\ref{eq:zmodequat}) we find
\be\nonumber
	\left(\dfrac{1-\gamma_5}{2}\right)\psi_0=\dfrac{e^{\mu t} \sqrt{\Pi}}{2\pi\sqrt{2}\rho}
	\left[\begin{array}{cc}
		0 & \sigma_\rho^\dag\sigma_2
		\end{array}
	\right]_{j\beta}
	\partial_\rho\left(\dfrac{\Phi}{\Pi}e^{-\mu t}\right).
\ee

We can now proceed to introduce this result into the Dirac equation, obtaining
\be
	&&\nonumber\left[\left(\begin{array}{cc}
		0 & -i\sigma_\rho^{\dag}\\
		i\sigma_\rho&0
		\end{array}
	\right)_{\alpha\beta}
	\left(\partial_{\mu}+\dfrac{1}{4}(\sigma_\mu^\dag\sigma_\nu-\sigma_\nu^\dag\sigma_\mu)\partial_\nu\ln\Pi\right)_{ij}-\mu\left(\begin{array}{cc}
		0 & -i\sigma_4^{\dag}\\
		i\sigma_4&0
		\end{array}
	\right)_{\alpha\beta}\mathcal{I}_{ij}\right]\cdot\\
	&&\cdot\dfrac{e^{\mu t} \sqrt{\Pi}}{2\pi\sqrt{2}\rho}
	\left(\begin{array}{cc}
		0 & \sigma_\rho^\dag\sigma_2
		\end{array}
	\right)_{j\beta}
	\partial_\rho\left(\dfrac{\Phi}{\Pi}e^{-\mu t}\right)	= 0
\ee
Using the relations (\ref{eq:sig1}-\ref{eq:sig2}) we obtain
\be\nonumber
	\left(\begin{array}{cc}
		0 & -i\sigma_\rho^{\dag}\\
		i\sigma_\rho&0
		\end{array}
	\right)_{\alpha\beta}\left(\begin{array}{cc}
		0 & \sigma_\rho^\dag\sigma_2
		\end{array}\right)_{j\beta}=\left(\begin{array}{cc}
		i\sigma_\rho^\dag\sigma_{\mu}\sigma_2 & 0
		\end{array}\right)_{j\alpha},
\ee
and
\be\nonumber
	\left(\begin{array}{cc}
		0 & -i\sigma_4^{\dag}\\
		i\sigma_4&0
		\end{array}
	\right)_{\alpha\beta}\left(\begin{array}{cc}
		0 & \sigma_\rho^\dag\sigma_2
		\end{array}\right)_{j\beta}=\left(\begin{array}{cc}
		\sigma_\rho^\dag\sigma_2 & 0
		\end{array}\right)_{j\alpha}.
\ee
From this equalities we get the Dirac equation in the following form
\be\label{eq:diraceq}
	&&\Bigg[\left(\partial_\mu+\dfrac{1}{4}(\sigma_{\mu}^\dag\sigma_\nu-\sigma_\nu^\dag\sigma_\mu)\partial_\nu\ln\Pi\right)\left(\dfrac{e^{\mu t} \sqrt{\Pi}}{2\pi\sqrt{2}\rho}i\sigma_\rho^\dag\partial_\rho\left(\dfrac{\Phi}{\Pi}e^{-\mu t}\right)\sigma_\mu\right)\nonumber\\
	&&-i\mu\dfrac{e^{\mu t} \sqrt{\Pi}}{2\pi\sqrt{2}\rho}\sigma_\rho^\dag\partial_\rho\left(\dfrac{\Phi}{\Pi}e^{-\mu t}\right)\Bigg]\sigma_2=0.
\ee

A straightforward calculation shows that the first term of the equation~(\ref{eq:diraceq}) becomes
\be\label{eq:diraceqt1}
	&&\left(\partial_\mu+\dfrac{1}{4}(\sigma_{\mu}^\dag\sigma_\nu-\sigma_\nu^\dag\sigma_\mu)\partial_\nu\ln\Pi\right)\left(e^{\mu t} \sqrt{\Pi}\ i\sigma_\rho^\dag\partial_\rho\left(\dfrac{\Phi}{\Pi}e^{-\mu t}\right)\sigma_\mu\right) =\nonumber\\
	&&2\mu e^{\mu t}\sqrt{\Pi}\partial_4\left(\Phi/\Pi e^{-\mu t}\right)+e^{\mu t}\dfrac{\partial_\mu\Pi}{\sqrt{\Pi}}\partial_\mu(\Phi/\Pi e^{-\mu t})+ e^{\mu t}\sqrt{\Pi}\Box(\Phi/\Pi e^{-\mu t})\nonumber\\
	&&-i\mu e^{\mu t}\sqrt{\Pi}\sigma_{\nu}\partial_\nu(\Phi/\Pi e^{-\mu t})-\sigma_\mu^\dag\dfrac{e^{\mu t}}{2}\dfrac{\partial_\mu\Pi}{\sqrt{\Pi}}\sigma_\nu\partial_\nu(\Phi/\Pi e^{-\mu t})\nonumber\\
	&&+\dfrac{\partial_\mu\Pi}{\sqrt{\Pi}}e^{\mu t}\partial_\mu(\Phi/\Pi e^{-\mu t})+\dfrac{1}{2}\sigma_\mu^\dag\dfrac{\partial_\mu\Pi}{\sqrt{\Pi}} e^{\mu t}\sigma_\nu\partial_\nu(\Phi/\Pi e^{-\mu t}),
\ee
while for the second term we have
\be\label{eq:diraceqt2}
	&&i\mu\ e^{\mu t}\sqrt{\Pi}\ \sigma_\nu^\dag\partial_\nu\left(\dfrac{\Phi}{\Pi}e^{-\mu t}\right)\nonumber\\
	&&\hspace{2cm}=i2\mu e^{\mu t}\sqrt{\Pi} \partial_4(\Phi/\Pi e^{-\mu t})+i\mu e^{\mu t}\sqrt{\Pi}\sigma_\nu\partial_\nu(\Phi/\Pi e^{-\mu t}).
\ee
Neglecting the normalization factor $(2\sqrt{2}\pi\rho)^{-1}$, the sum of the two terms yields:
\be
	\dfrac{1}{\sqrt{\Pi}}\left(2\partial_{\mu}\Pi\ \partial_\mu\left(\dfrac{\Phi}{\Pi}e^{-\mu t}\right)+\Pi\ \Box\left(\dfrac{\Phi}{\Pi}e^{-\mu t}\right)\right)=0.
\ee
Performing the derivatives and using that $\Box\Pi=0$, we finally obtain:
\be\label{eq:KGtmu}
	\sqrt{\Pi}^{-1}\left(\Box\Phi-2\mu\partial_4\Phi+\mu^2 \Phi\right)=0.
\ee

Substituting the expression of $\Phi(r,t)$ as given in~(\ref{eq:phi}) we immediately see that equation (\ref{eq:KGtmu}) is satisfied, which concludes our proof. 

It is interesting to notice that~(\ref{eq:KGtmu}) corresponds to the massless Klein-Gordon equation at finite temperature and chemical potential. We also observe that  (\ref{eq:phi}) is nothing else but the Klein-Gordon propagator for anti-periodic boundary condition in the $t$ direction, being such that $\Phi(t)=-\Phi(t+1/T)$. On the other hand, the Klein-Gordon propagator with periodic boundary conditions is given by
\be
	\Phi_S(r,t)=\left(\Pi(r,t)-1\right)\left(\cos(r \mu)+\dfrac{\sin(\pi t T)\sin(r \mu)}{\sinh(\pi r T)}\right),
\ee
and considering the limit for vanishing chemical potential we obtain nothing else than the $\Pi(r,t)$ used in the instanton calculus:
\be
	\lim_{\mu\rightarrow 0}\Phi_S(r,t)=1+\dfrac{\pi\rho^2}{T r}\dfrac{\sinh(2\pi r T)}{\cosh(2\pi r T)-\cos(2\pi t T)} =\Pi(r,t)
\ee

We thus see that both the instanton profile and the associated fermionic zero mode, are written in terms of propagators of Klein-Gordon scalar particles with appropriate boundary conditions.

\section{Phase diagram of the intanton liquid}\label{sec:phasediagram}

The knowledge of the analytic form of the zero mode opens the possibility to explore the full phase diagram of the instanton model. In this section we use the result obtained in the previous section and derive the partition function of the model showing that this can be numerically evaluated also at finite temperature and chemical potential.

Next, we will present some results on the phase diagram of the model, and in particular the dependence of the instanton density from the temperature and quark chemical potential. This knowledge is important not only to understand the structure of the QCD vacuum, but also in connection to chiral symmetry breaking, since the instanton density appears in the mean field relation:
\be\label{eq:mfcc}
	\langle q\bar{q}\rangle=-\dfrac{1}{\pi\rho}\left(\dfrac{3}{2}n\right)^{1/2}.
\ee
As a result, we will see that, as $T$ and $\mu$ grow, instantons tend to organize in molecules and the density $n$ appearing in the formula is not the total density of instanton, but only the fraction of instanton that does not coalesce.

Let us stress from the outset that the study performed in this section is for illustrative purposes and no conclusion on the phase transitions in the full QCD phase diagram will be attempted. On the other hand these results can be seen as a first attempt to study the effects related to the topological structure of the QCD vacuum  on the phase diagram.

\subsection{Instanton partition function and overlap matrix elements}\label{sec:partfunctia}
In the instanton model the quantum field theory describing the QCD vacuum is reduced to an effective theory where a finite number of effective degrees of freedom, the collective coordinates of the instantons $\Omega_i=\{z_i,\rho_i,U_i\}$ (position, size and color orientation), are treated in a statistical mechanical framework. In this context the partition function of the instanton ensemble is given by
\be\label{eq:partfunc}
	Z=\sum_{N_+N_-}\frac{1}{N_+!N_-!}\int\prod_i^{N_+N_-}[\textrm{d}\Omega_in(\rho_i)\rho_i^{N_f}]\exp(-S_{int})\prod_f^{N_f}\det(i\sc{D}+im_f-i\mu\gamma_4),
\ee
where the sum runs over the total number of  instantons $(N_+)$ and anti-instantons $(N_-)$. The single-instanton amplitude $n(\rho_i)$ contains the semiclassical tunneling rate (including one-loop quantum corrections), as well as the Jacobian arising from the introduction of collective coordinates~\cite{'tHooft:1976fv}. The instanton interaction can be divided into a gluonic part $S_{int}$ and a fermionic part represented by the determinant of the Dirac operator of the instanton liquid model. 

The fermionic determinant, in the chiral limit, can be calculated in the subspace of zero modes; by neglecting higher order corrections, in particular
\be
	\det(i\sc{D}-i\mu\gamma_4)\approx\det\left(\begin{array}{cc}
		0 & T_{IA}(\mu,T)\\
		T_{AI}(\mu,T)&0
		\end{array}\right),
\ee
where the $T_{IA}$ matrix elements are given by the overlap integrals
\be\label{eq:tiadef}
	T_{IA}=\int_0^{1/T}\textrm{d}\tau\int\textrm{d}\vec{x}\psi_I^\dag(\tau-z_4,\vec{x}-\vec{z};\mu,T)(i\sc{D}-i\mu\gamma_4)\psi_A(\tau,\vec{x};\mu,T).
\ee
The nonhermiticity of the finite-$\mu$ Dirac operator is reflected by the fact that
\be
	T_{AI}(\mu,T)=T_{IA}^\dag(-\mu)\neq T^\dag_{IA}(\mu,T),
\ee
i.e., the fermionic determinant is complex, entailing the well-known ``sign"-problem in the partition function.

The overlap matrix element depends on the distance $z_\nu$ between instantons, the instanton sizes $\rho_{I,A}$ and the relative color orientation $U=U_I^\dag U_A$, Which can be characterized using the four-vector $u_\nu=1/(2i)\textrm{Tr}(U\sigma_\nu)$. The definite chirality of the zero modes (in the limit of vanishing current quark masses) entails that the~$I-I$ and $A-A$ matrix elements are zero. In vacuum, Lorentz invariance implies that the overlap matrix element can be characterized by a single scalar function \cite{Shuryak:1991jg,Khoze:1990nt}, e.g. $T_{IA}=iu\cdot\hat{z}f(z)$. At $T,\mu\neq0$ this is no longer true, and the matrix elements have the more general structure
\be\label{eq:tiadec}
	T_{IA}=iu_4f_1+i(\vec{u}\cdot\hat{z})f_2
\ee
with $f_i=f_i(|\vec{z}|,z_4,\rho_I,rho_A)$ and $\hat{z}=\vec{z}/|\vec{z}|$. These functions can be calculated by writing the zero mode solutions, derived in the precedent section, in the form $\psi_i^a=\phi_\nu(x)(\gamma_\nu)_{ij}U^{ab}\chi_j^b$. In the general case $\mu,T\neq0$, we have 
\be
	\phi_\nu(x)=\delta_{\nu 4}\phi_4(r,x_4)+\delta_{\nu i}\hat{r}_i\phi_r(r,x_4).
\ee
and in terms of $\phi_{nu}$, the overlap matrix elements are given by
\be
	T_{IA}&=&2iu_\alpha\int_0^{1/T}\textrm{d}\tau\int\textrm{d}\vec{x}(\phi_\nu^A\partial_\nu\phi^I_\alpha-\phi^A_\nu\partial_\alpha\phi^I_\nu+\phi^A_\alpha\partial_\nu\phi^I_\nu\nonumber\\
	&&-\mu(\phi_4^A\phi^I_\alpha-\delta_{\alpha 4}\phi_\nu^A\phi_\nu^I+\phi_\alpha^A\phi^I_4)
\ee 
where $\phi_\nu^A=\phi_\nu(x-z)$ and $\phi_\nu^I=\phi_\nu(x)$. Comparing (\ref{eq:tiadec}) with the above equation we find that the functions $f_{1,2}$ can be expressed by the integrals:
\be\label{eq:f1}
	f_1&=&2\int_0^{1/T}\textrm{d}\tau\int\textrm{d}\vec{x}\Bigg(\phi_4^A(\partial_4\phi_4^I+2\phi_r^I/r+\partial_r\phi_r^I)\nonumber\\
	&&+\frac{\vec{r}-\vec{z}}{|r-z|}\cdot\hat{r}\ \phi_r^A(\partial_r\phi_4^I-\partial_4\phi^I_r)-\mu\left(\phi_4^A\phi_4^I-\frac{\vec{r}-\vec{z}}{|r-z|}\cdot\hat{r}\ \phi^A_r\phi^I_r\right)\Bigg),\\\label{eq:f2}
	f_2&=&2\int_0^{1/T}\textrm{d}\tau\int\textrm{d}\vec{x}\Bigg(\hat{z}\cdot\frac{\vec{r}-\vec{z}}{|r-z|}\ \phi_r^A(\partial_4\phi_4^I+2\phi_r^I/r+\partial_r\phi_r^I)\nonumber\\
	&&+\hat{r}\cdot\hat{z}\phi_4^A(\partial_4\phi_r^I-\partial_r\phi_4^I)-\mu\left(\hat{z}\cdot{r}\phi^A_4\phi^I_r+\hat{z}\cdot\frac{\vec{r}-\vec{z}}{|r-z|}\phi_r^A\phi_4^I\right)\Bigg).
\ee

\begin{figure}
       \centering
        \subfigure{\includegraphics[width=.7\textwidth]{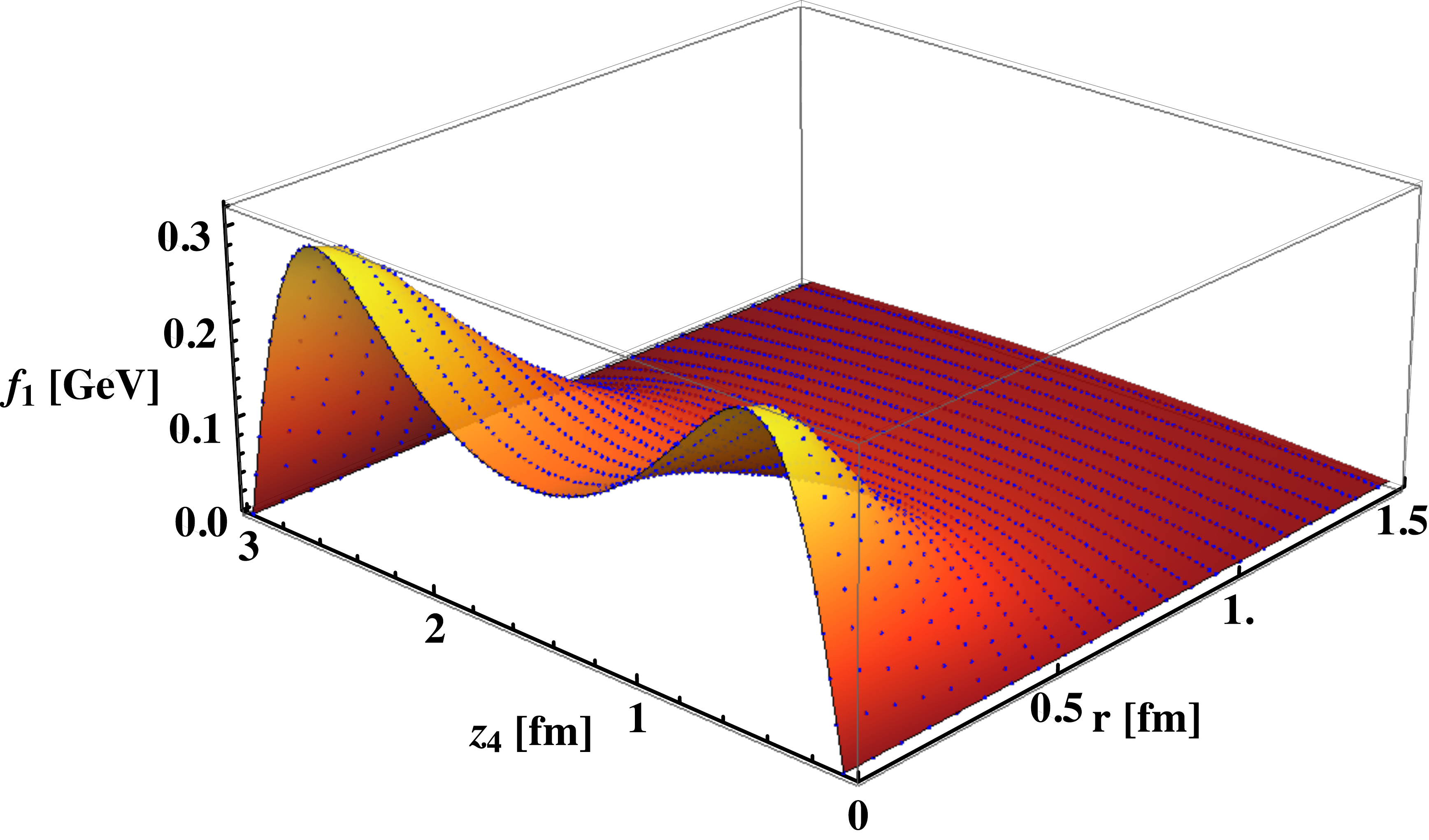}}\\
       \subfigure{\includegraphics[width=.7\textwidth]{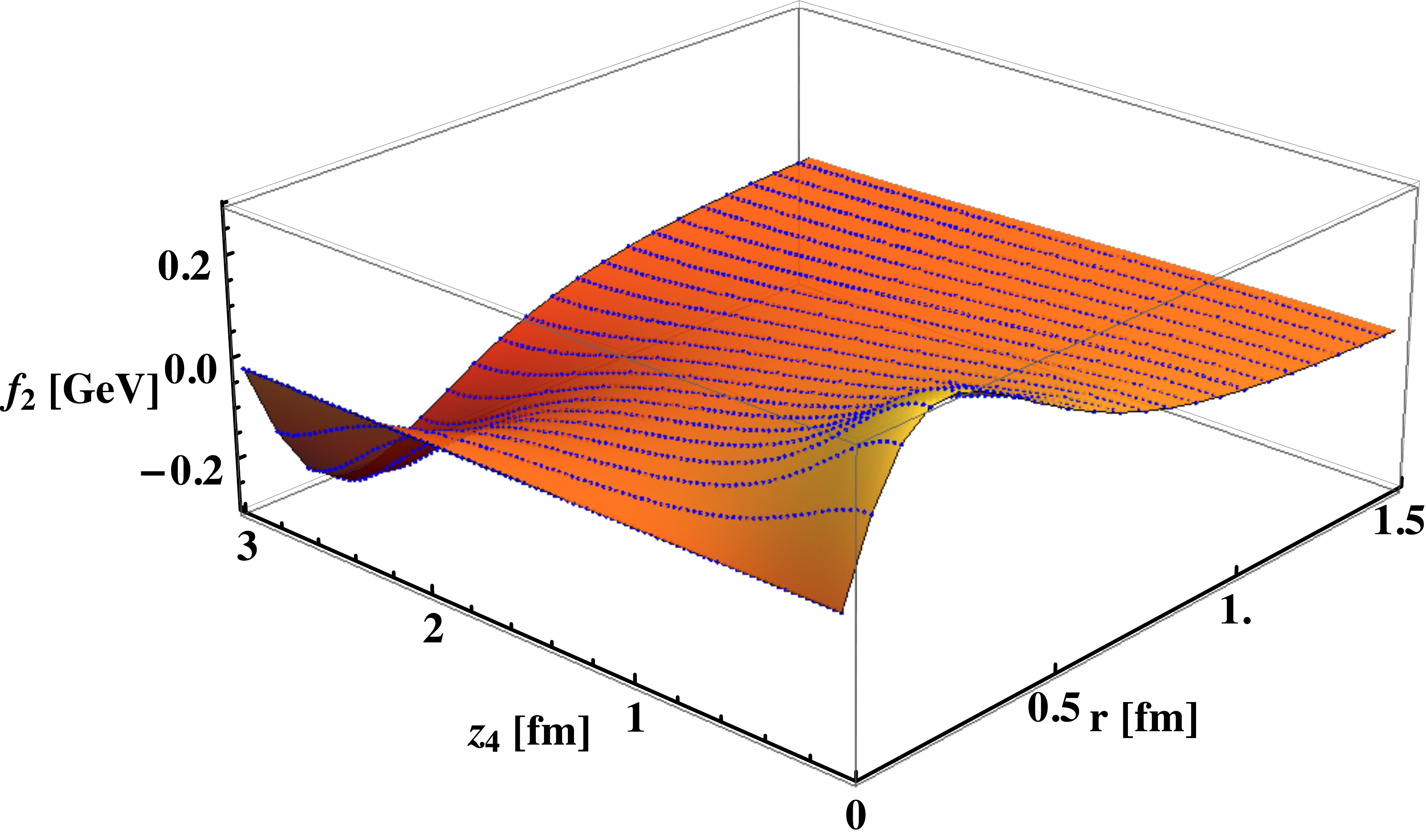}}
\caption{$f_1$ (upper panel) and  $f_2$ (lower panel) functions evaluated as a function of $z_4$ and $r$ on a grid (blue points). The solid surface, obtained interpolating the points, is plotted to show that the grid is sufficiently dense to reproduce smoothly the details of the functions.}
\label{fig:f1f2}
\end{figure}

For the subsequent study of the thermodynamics of the model we need to integrate the $f_{1,2}$ functions in the $z_4$ and $r$ coordinates, which means that for each value of temperature and chemical potential we need to have a sufficient number of points in the $(z_4,r)$ plane to ensure integral convergence implying in turn large computational times. As an example we show in Fig.~\ref{fig:f1f2} the functions $f_{1,2}$ evaluated at $T=65$ Mev and $\mu=0$ MeV. 

\subsection{Chiral phase transition at finite temperature and chemical potential}\label{sec:chiraltrans}
In the instanton liquid model the study of QCD thermodynamics is reduced to a statistical mechanics treatment of an ensemble of classical pseudoparticles. In this context spontaneous chiral symmetry breaking in the vacuum is generated by randomly distributed uncorrelated instantons--anti-instantons which allow for a delocalization of the associated quark quasi-zero modes corresponding to the formation of a nonzero $\langle q\bar{q}\rangle$ condensate state (see, e.g.,~\cite{Schafer:1996wv} for a review of the argument). This can be understood, by thinking in terms of quarks that can travel arbitrarily long distances by randomly jumping from one instanton to another (anti-instanton) and thus may carry their chiral charge to spatial infinity. 

In this picture chiral restoration can in principle proceed in two ways: either instantons disappear altogether or they rearrange into some finite clusters which no longer support any finite $\langle q\bar{q}\rangle$ condensate.

From the most recent calculations of the most refined version of the instanton model at finite temperature~\cite{Wantz:2009mi}, it emerges that instantons survives well beyond $T_c$, and the same looks plausible also for the case of finite chemical potential. This suggest that chiral restoration will appear as the consequence of the rearrangement of single instantons into I-A molecules~\cite{Ilgenfritz:1988dh,Ilgenfritz:1994nt,Schafer:1994nv}. 

The investigation of the interplay between the different components in the instanton partition function at finite temperature and chemical potential can be addressed, by adopting a simplified version of the instanton model, the so called ``cocktail model"~\cite{Ilgenfritz:1988dh,Rapp:1999qa}. In this model, which we follow closely, the ensemble is decomposed into a mixture of random (``atomic") and ``molecular" configurations, which yields a grand canonical partition function of the form
\be
	\mathcal{Z}^{a+m}=\sum_{N_a,N_m}\dfrac{(z_aV_4)^{N_a}}{N_a!}\dfrac{(z_mV_4)^{N_m}}{N_m!}.
\ee
In the thermodynamic limit $V_4\rightarrow\infty$, and using the Stirling formula, the thermodynamic potential becomes
\be
	\Omega^{a+m}(n_a,n_m;T,\mu)=-\dfrac{\ln[\mathcal{Z}^{a+m}]}{V_4}=-n_a\ln\left[\dfrac{e z_a}{n_a}\right]-n_a\ln\left[\dfrac{e z_m}{n_m}\right].
\ee
The atomic and molecular ``activities" are
\be
	z_a&=&2 C\rho^{b-4}e^{-S_{int}}d_{pert}(T)\langle T_{IA}(T,\mu)T_{AI}(T,\mu)\rangle^{N_f/2}\nonumber\\
	&=&2 C\rho^{b-4}e^{-S_{int}}d_{pert}(T)\left(\dfrac{n_a}{2}\int\textrm{d}^4z\textrm{d}U[T_{IA}(T,\mu)T_{AI}(T,\mu)]\rho^2\right)^{N_f/2}\\
	z_m&=&C^2\rho^{2(b-4)}e^{-2S_{int}}d_{pert}(T)^2\langle [T_{IA}(T,\mu)T_{AI}(T,\mu)]^{N_f}\rangle\nonumber\\
	&=&C^2\rho^{2(b-4)}e^{-2S_{int}}d_{pert}(T)^2\int\textrm{d}^4z\textrm{d}U[T_{IA}(T,\mu)T_{AI}(T,\mu)]^{N_f}\rho^{2N_f}.
\ee
where $d_{pert}(T)$ is a perturbative suppression factor which has the form
\be
	d_{pert}(T)&=&\exp\Big(-\dfrac{2N_c+N_f}{3}(\pi\rho T)^2-\left(1+\dfrac{N_c}{6}-\dfrac{N_f}{6}\right)\cdot\nonumber\\
		&&\hspace{5mm}(-\ln(1+(\pi\rho T)^2/3)+0.15/(1+0.15(\pi\rho T)^{-3/2})^8)\Big).
\ee

The underlying approximation in this approach is that the values of the hopping amplitude $T_{IA}$ in each individual configuration are replaced by a product of their mean square values in an uncorrelated ensemble. Another important approximation introduced in the model is the reduction of the gluonic interaction to an average repulsion term
\be
	S_{int}=\kappa\rho^4(n_a+2n_m),\quad \kappa=\beta/(2\bar{\rho}(N/V))\nonumber\\
	\beta=b/2+3N_f/4-2,\quad b=(11/3)N_c-(2/3)N_f.
\ee
The free parameter $\kappa$ characterizes the diluteness of the ensemble. In the case under consideration with $N_c=3$ and $N_f=2$ the value which reproduces the phenomenological diluteness of the instanton vacuum is $\kappa\approx 130$~\cite{Rapp:1999qa}.

Within this assumption the color dependence in the activities only enters through the combination of the $T_{IA}$'s which then can be integrated analytically, rendering the fermionic determinant real. For two flavor one obtains
\be
	z_a(z_4,r)&\propto&\int\textrm{d}U\ T_{IA}(T,\mu)T^\dag_{IA}(T,-\mu)\nonumber\\
	&=&\dfrac{1}{2N_c}(f_1^+f_1^-+f_2^+f_2^-)\nonumber\\
	z_m(z_4,r)&\propto&\int\textrm{d}U\ (T_{IA}(T,\mu)T^\dag_{IA}(T,-\mu))^{N_f}\nonumber\\
	&=&\dfrac{(2N_c-1)(f_1^+f_1^-+f_2^+f_2^-)^2+(f_1^+f_2^--f_1^-f_2^+)^2}{4N_c(N_c^2-1)},\nonumber
\ee
where $f_i^\pm\equiv f_i(\pm\mu)$ are the functions defined in the previous section.
The last free parameter of the model is the constant $C$ that we fixed imposing that the absolute minimum of the thermodynamic potential $\Omega(T=0,\mu=0;n_a,n_m)$ appears at a total instanton density of $N/V=n_a+2n_m=1.0\ \textrm{fm}^{-4}$ which is the phenomenological instanton density in the vacuum. 

\begin{figure}
       \centering
        \subfigure{\includegraphics[width=.7\textwidth]{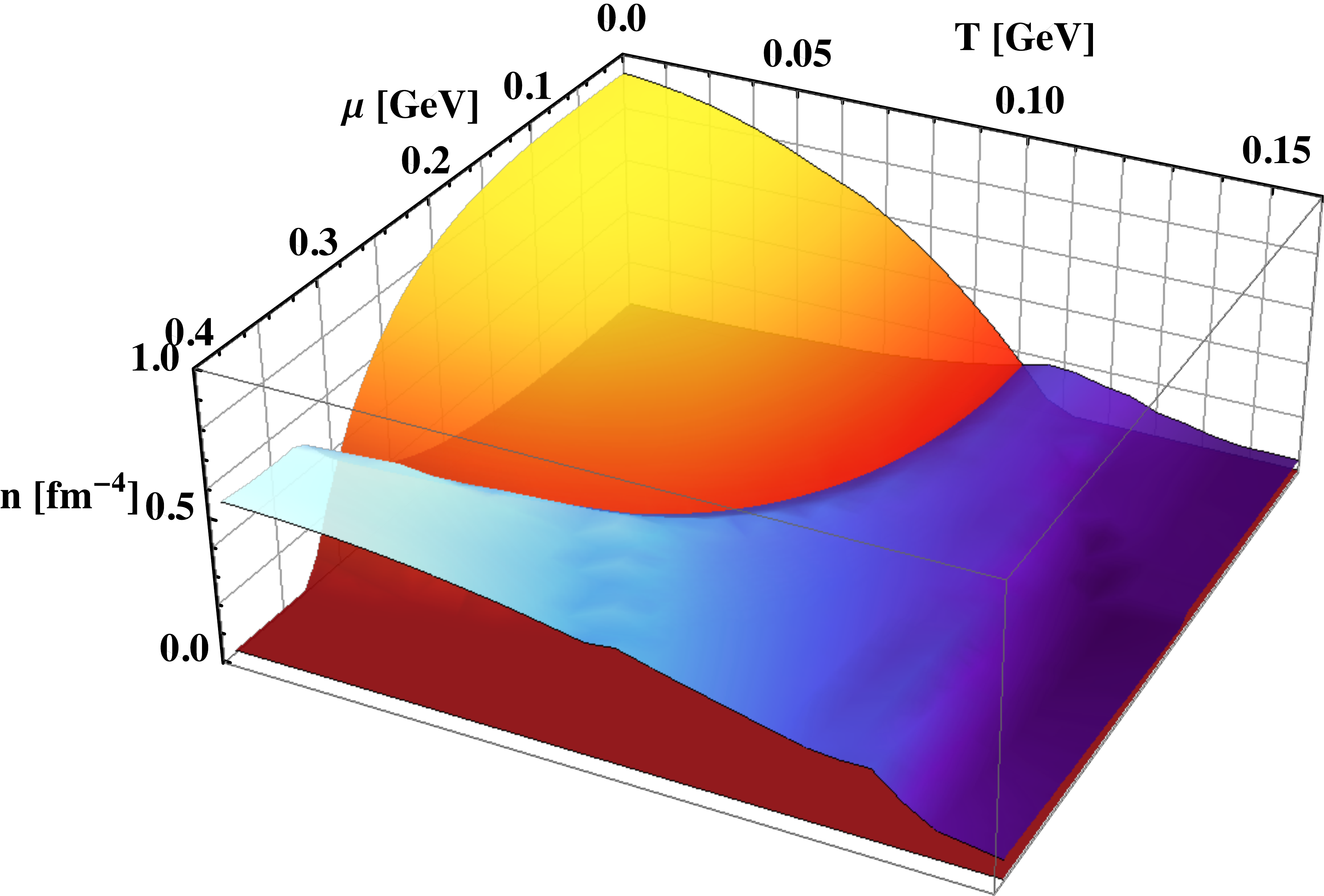}}\\
       \subfigure{\includegraphics[width=.7\textwidth]{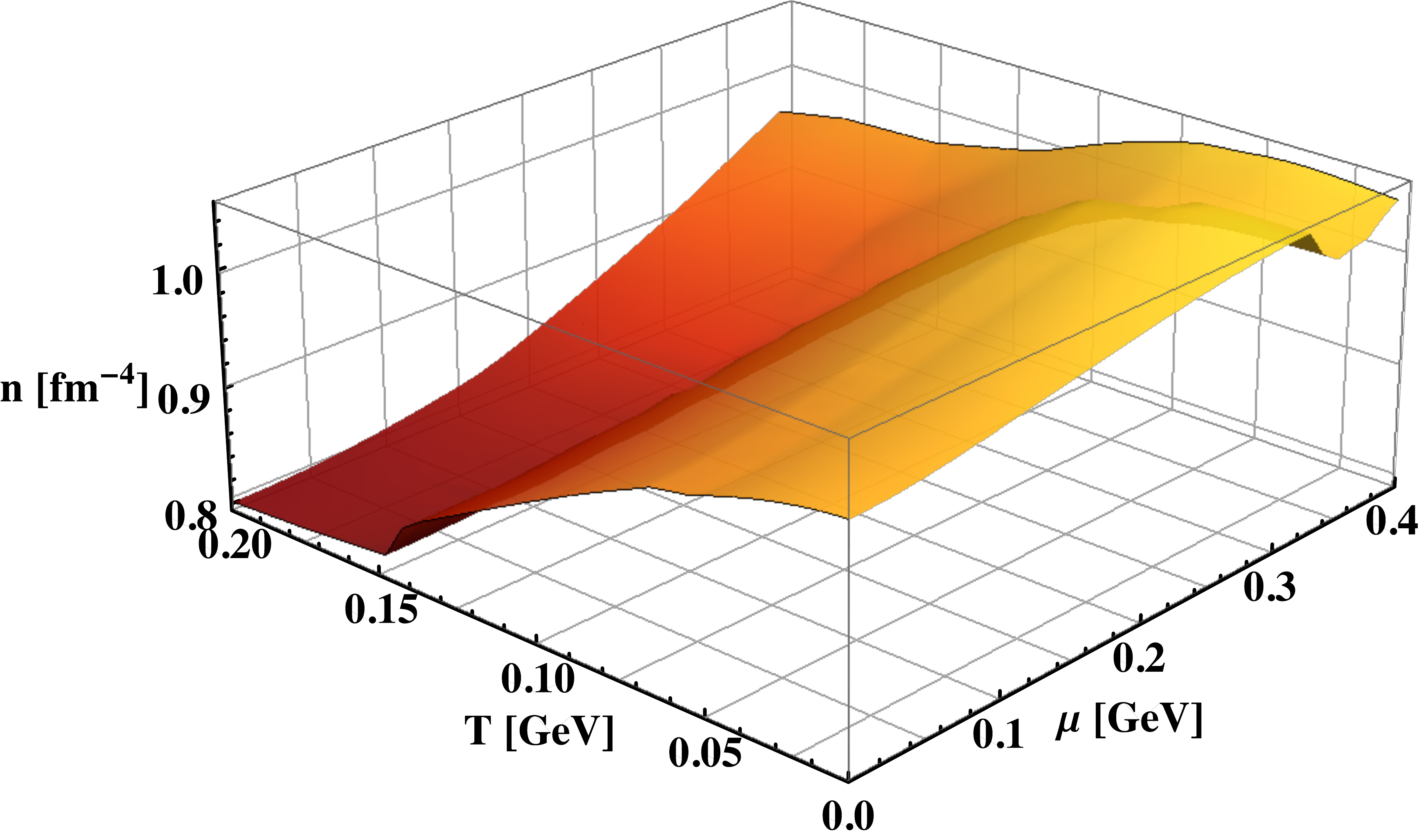}}
\caption{Upper panel: Instanton density (red-yellow colors) and molecule density (blue colors) as a function of the temperature and chemical potential. Lower panel: total density (instantons + molecules).}
\label{fig:instdens}
\end{figure}

The numerical results obtained for the $N_c=3$, $N_f=2$ cocktail model in the chiral limit, are presented in Fig.~\ref{fig:instdens}. The atomic density decreases as expected with growing temperature and chemical potential, while on the contrary the molecular component of the ensemble becomes more and more important. The behavior of the total density is represented in the upper panel: at small chemical potential when the temperature grows the instantons tend to diminish and eventually disappear at a temperature around $T\approx 150$~MeV. On the other hand, the presence of a non-vanishing density of quarks tend to preserve instantons.

The mean field approximation allows establishing the connection between the density of the atomic component and the chiral quark condensate shown in (\ref{eq:mfcc}). From this latter relation one can connect restoration of chiral symmetry with the disappearance of atomic instantons. Indeed, our results, presented in Fig.~\ref{fig:instdens}, show a phase transition at $\mu\approx 320$~MeV for vanishing temperature and at $T\approx 120$ MeV at zero chemical potential. This numbers agrees with previous calculation in the instanton model~\cite{Schafer:1995pz,Rapp:1999qa,Schafer:1998up,Wantz:2009mi} and confirm that the chiral transition appears when the total density of instantons is considerably different from zero in the full phase diagram. 

\section{Conclusions}\label{sec:conclusion}
The instanton model has been extensively used in the study of the non-perturbative sector of QCD. The interest in it lies in the early discovery of the Dirac zero mode that the presence of instantons in the ground state of the theory gives rise to. Despite all the development in the field, the general form of the zero mode profile in the general case of  $T\neq0$ and $\mu\neq0$ has been a pending issue.

In this paper we fill this gap presenting the zero mode solution of the Dirac equation at finite temperature and chemical potential in the background of a singular gauge instanton. Starting from an Ansatz inspired by the known solutions in the limiting cases $T\neq0$, $\mu=0$ and $T=0$, $\mu\neq0$ we show that the proposed profile indeed satisfies the Dirac equation.

The analytic expression for the zero-mode presented in this paper represents the starting point for the investigation of the role of instantons in the full phase diagram of QCD. In particular, it would be interesting to understand the contribution of such gauge configuration to the restoration of the chiral symmetry breaking around the critical point; on the other hand, also the $U(1)$ axial anomaly and its possible restoration, can now be studied in the general setting of finite temperature and chemical potential. 

As a first step in this direction, we used the ``cocktail" version of the model in order to investigate the instanton density in the phase diagram. In this framework, the analysis of the balance between the atomic and molecular components of the ensemble provides some clues for unveiling the mechanism responsible for chiral symmetry restoration. In particular, we have shown that the restoration happens when the total density of instantons is still finite and, as expected, connected with the formation of instanton molecules in the vacuum.

By studying the instanton model with more sophisticated techniques than the simple ``cocktail" approach used here (e.g. the Interacting Instanton Liquid model developed in~\cite{Schafer:1995pz}), we can hope to extract more information on the chiral dynamics induced by instantons in the QCD phase diagram.

\acknowledgments
I would like thank D. Binosi and P. Faccioli for their help and support. Numerical calculations were performed using the Aurora supercomputer at FBK/Trento~\endnote{http://web.infn.it/aurorascience/}.
This research is supported by the AuroraScience project, which is funded jointly by the Provincia Autonoma di Trento (PAT)  and the Istituto Nazionale di Fisica Nucleare (INFN).

\appendix

\section{Properies of $\eta$ symbol}\label{app:thoofts}
The 't-Hooft $\eta$-symbols entering in the analytic expression of the instanton profile, are defined through~\cite{'tHooft:1976fv}
\be
	&&\eta_{a\mu\nu}=\varepsilon_{a\mu\nu}+\delta_{a\mu}\delta_{\nu 4}-\delta_{a\nu}\delta_{\mu 4},\nonumber\\
	&&\bar{\eta}_{a\mu\nu}=\varepsilon_{a\mu\nu}-\delta_{a\mu}\delta_{\nu 4}+\delta_{a\nu}\delta_{\mu 4}.\nonumber
\ee
The $\eta$-symbols are (anti) self-dual in the vector indices
\be
	\eta_{a\mu\nu}=\frac{1}{2}\varepsilon_{\mu\nu\alpha\beta}\eta_{a\alpha\beta},&\ \bar{\eta}_{a\mu\nu}=-\frac{1}{2}\varepsilon_{\mu\nu\alpha\beta}\bar{\eta}_{a\alpha\beta}, &\ \eta_{a\mu\nu}=-\eta_{a\nu\mu}\nonumber
\ee

Contraction involving the  $\eta$-symbols can be done using the following relations:
\be
	&&\eta_{a\mu\nu}\eta_{b\mu\nu}=4\delta_{ab},\nonumber\\
	&&\eta_{a\mu\nu}\eta_{a\mu\rho}=3\delta_{\nu\rho},\nonumber\\
	&&\eta_{a\mu\nu}\eta_{a\mu\nu}=12,\nonumber\\
	&&\eta_{a\mu\nu}\eta_{a\rho\lambda}=\delta_{\mu\rho}\delta_{\nu\lambda}-\delta_{\mu\lambda}\delta_{\nu\rho}+\varepsilon_{\mu\nu\rho\lambda},\nonumber\\
	&&\eta_{a\mu\nu}\eta_{b\mu\rho}=\delta_{ab}\delta_{\nu\rho}+\varepsilon_{abc}\eta_{c\nu\rho},\nonumber\\
	&&\eta_{a\mu\nu}\bar{\eta}_{b\mu\nu}=0,\nonumber\\
	&&\varepsilon_{abc}\eta_{b\mu\nu}\eta_{c\rho\lambda}=\delta_{\mu\rho}\eta_{a\nu\lambda}-\delta_{\mu\lambda}\eta_{a\nu\rho}+\delta_{\nu\lambda}\eta_{a\mu\rho}-\delta_{\nu\rho}\eta_{a\mu\lambda},\nonumber\\
	&&\varepsilon_{\lambda\mu\nu\sigma}\eta_{a\rho\sigma}=\delta_{\rho\lambda}\eta_{a\mu\nu}+\delta_{\rho\nu}\eta_{a\lambda\mu}+\delta_{\rho\mu}\eta_{a\nu\lambda},\nonumber
\ee
with the same relations holding for $\bar{\eta}_{a\mu\nu}$, except for
\be
	&&\bar{\eta}_{a\mu\nu}\bar{\eta}_{a\rho\lambda}=\delta_{\mu\rho}\delta_{\nu\lambda}-\delta_{\mu\lambda}\delta_{\nu\rho}-\varepsilon_{\mu\nu\rho\lambda}.\nonumber
\ee

\section{Gamma and sigma matrices}\label{app:notation}
Let us define
\be
	\gamma_4^E=\gamma_4^M, &\  \gamma_m^E=-\gamma_m^M, &\  m=1,2,3.
\ee
where the explicit form of the gamma matrices is given by
\begin{displaymath}
\gamma_{\mu}=
	\left(\begin{array}{cc}
		0 & -i\sigma_{\mu}^{\dag}\\
		i\sigma_{\mu}&0
		\end{array}
	\right),
\end{displaymath}
with $\sigma_{\mu}=(\vec{\sigma},-i)$ and $\sigma$'s the usual Pauli matrices with $\sigma^a\sigma^b=\delta^{ab}+i\epsilon^{abc}\sigma^c$.
The  resulting $\gamma_5$ will be
\begin{displaymath}
\gamma_5\equiv\gamma_1\gamma_2\gamma_3\gamma_4=
	\left(\begin{array}{cc}
		1 & 0\\
		0&-1
		\end{array}
	\right).
\end{displaymath}

Directly from the definition we have 
\be\label{eq:sig1}
	\sigma_{\mu}^{\dag}\sigma_{\nu}+\sigma_{\nu}^{\dag}\sigma_{\mu}=\sigma_{\mu}\sigma_{\nu}^{\dag}+\sigma_{\nu}\sigma_{\mu}^{\dag}=2\delta_{\mu\nu}.
\ee	
so that
\be\label{eq:sig2}
	\sigma_\mu^\dag\sigma_\nu\sigma_\mu^\dag=-2\sigma_\nu^\dag,&\qquad & \sigma_\mu\sigma_\nu^\dag\sigma_\mu=-2\sigma_\nu.
\ee
Notice finally that the 't Hooft $\eta$ symbols can be defined using the $\sigma_{\mu}$ by
\be
	i\eta^a_{\mu\nu}\sigma^a=\dfrac{1}{2}(\sigma_{\mu}\sigma_{\nu}^{\dag}-\sigma_{\nu}\sigma_{\mu}^{\dag})\\
	i\bar{\eta}^a_{\mu\nu}\sigma^a=\dfrac{1}{2}(\sigma_{\mu}^{\dag}\sigma_{\nu}-\sigma_{\nu}^{\dag}\sigma_{\mu}),
\ee

\bibliography{zmodes}

\end{document}